\documentclass[aps,prl,reprint,a4paper,showpacs,superscriptaddress]{revtex4-1}
\usepackage{amsmath}
\usepackage[latin9]{inputenc}
\usepackage{graphicx}
\usepackage{hyperref}
\usepackage{siunitx}
\usepackage{graphicx}
\usepackage[normalem]{ulem}
\usepackage{color}
\DeclareSIUnit\gauss{G}

\begin{document}
\title{Sensitivity of ultracold atoms to quantized flux in a superconducting ring}

\author{P. Weiss}
\email{pweiss@pit.physik.uni-tuebingen.de}
\author{M. Knufinke}
\affiliation{CQ Center for Collective Quantum Phenomena and their
Applications in LISA$^+$, Physikalisches Institut, Eberhard Karls Universit\"at
T\"ubingen, Auf der Morgenstelle 14, D-72076 T\"ubingen,
Germany}
\author{S. Bernon}
\affiliation{CQ Center for Collective Quantum Phenomena and their
Applications in LISA$^+$, Physikalisches Institut, Eberhard Karls Universit\"at
T\"ubingen, Auf der Morgenstelle 14, D-72076 T\"ubingen,
Germany}
\affiliation{Present address: Laboratoire Photonique, Num\'erique et Nanosciences-LP2N Universit\'e Bordeaux - IOGS - CNRS: UMR 5298 - Rue Fran\c{c}ois Mitterrand, 33400 Talence, France}
\author{D. Bothner}
\author{L. S\'ark\'any}
\author{C. Zimmermann}
\author{R. Kleiner}
\author{D. Koelle}
\author{J. Fort\'agh}
\email{fortagh@uni-tuebingen.de}
\author{H. Hattermann}
\affiliation{CQ Center for Collective Quantum Phenomena and their
Applications in LISA$^+$, Physikalisches Institut, Eberhard Karls Universit\"at
T\"ubingen, Auf der Morgenstelle 14, D-72076 T\"ubingen,
Germany}
\begin{abstract}
We report on the magnetic trapping of an ultracold ensemble of $^{87}$Rb atoms close to a superconducting ring prepared in different states of quantized magnetic flux. 
The niobium ring of \SI{10}{\micro m} radius is prepared in a flux state $n \Phi_0$, with $\Phi_0 = h / 2e$ the flux quantum and $n$ varies between $\pm 5$.
An atomic cloud of \SI{250}{nK} temperature is positioned with a harmonic magnetic trapping potential at $\sim$\SI{18}{\micro m} distance below the ring. 
The inhomogeneous magnetic field of the supercurrent in the ring contributes to the magnetic trapping potential of the cloud. 
The induced deformation of the magnetic trap impacts the shape of the cloud, the number of trapped atoms as well as the center-of-mass oscillation frequency of Bose-Einstein condensates. When the field applied during cooldown of the chip is varied, the change of these properties shows discrete steps that quantitatively match flux quantization.
\end{abstract}
\maketitle
The coherent coupling between atoms and single flux quanta in a superconducting circuit is an important ingredient of future cold atom-superconductor hybrid quantum systems in which quantum states are transferred from one system to the other.
The construction of such a hybrid quantum system is targeted in a number of recent experiments and proposals \cite{Rabl2006, Petrosyan2008, Petrosyan2009, Verdu2009, Bernon2013, Patton2013, Hogan2012, Kim2011Wellstood, Wallquist2009, Hafezi2012, jessen2013, Minniberger2014, Pritchard2014}, and should allow the study of fundamental interactions between the two systems \cite{Scheel05, Skagerstam06, Cano08a, Kasch2010, Sapina2013, Naides2014}.
One of the most prominent consequences of the existence of a macroscopic wave function in superconductors is the quantization of the magnetic fluxoid, which has been shown e.g. in superconducting rings and cylinders \cite{doll61, deaver61}.
In superconducting atom chip experiments, trapped  Abrikosov vortices have been used to magnetically trap atoms in spatially inhomogeneous fields \cite{Shimizu2009, Emmert09, Mueller2010a, Mueller2010, Zhang2012, Siercke2012}. 
These traps are affected by the motion of the vortices that potentially cause heating and losses of the cold atoms \cite{Scheel2007, Nogues2009}. 
Pinning the vortices would suppress this noise source and could be used to generate subwavelength magnetic lattices \cite{Romero2013}, as well as hybrid quantum systems based on atom traps formed by single pinned flux quanta \cite{Sokolovsky2014}.
The creation of a flux superposition state could give rise to a superposition of the magnetic trapping potential and therefore of the position of an atomic ensemble \cite{Singh09}. Ultrafast coupling (10 ns) between cold Rydberg atoms and SQUIDs has also been theoretically predicted \cite{Patton2013a}. 
It is therefore crucial to understand the impact of single flux quanta onto an ensemble of trapped ultracold atoms.

In this Letter, we report on how a discrete number of flux quanta stored in a superconducting ring affects the trapping parameters of a superimposed magnetic trap.
The discrete nature of the magnetic flux in the ring is observed both in the atom number and the oscillation frequency of atoms in the trap.

\begin{figure}
\centerline{\includegraphics[width=.48\textwidth]{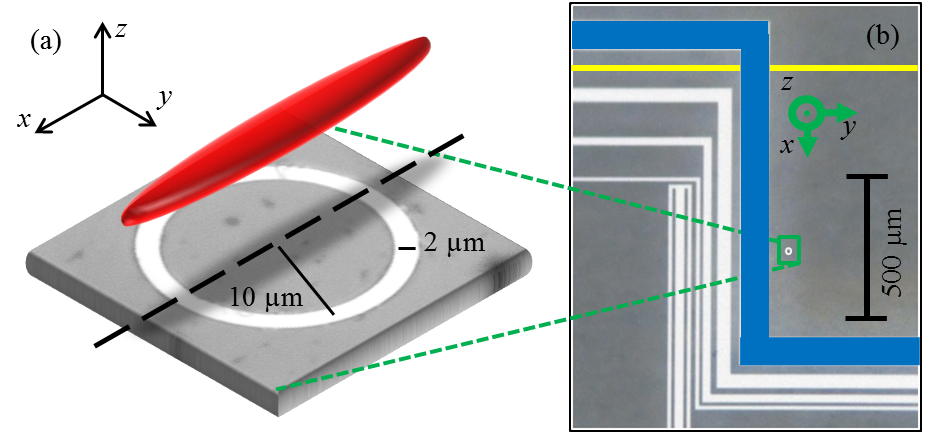}}
\caption{(a) Sketch of an atomic ensemble trapped at the superconducting ring. 
(b) Optical image of parts of the superconducting chip.  Shown are four trapping wires, the \SI{100}{\micro\meter} wide trapping wire used for the present experiment is highlighted in blue. The location of the confinement wire is sketched in yellow. The center of the superconducting ring is located \SI{70}{\micro\meter} from the right edge of the trapping wire. The chip is mounted upside down on the cryostat, so gravity points in $+z$ direction.}
\label{fig:chip}
\end{figure}
We magnetically trap an ensemble of cold $^{87}$Rb atoms on a superconducting atom chip and guide it to the vicinity of a superconducting ring, as sketched in Fig.\ \ref{fig:chip}(a).
The atom chip (Fig. \ref{fig:chip}(b)) is a sapphire substrate ($\sim$\SI{330}{\micro \meter} thick) with patterned niobium thin film structures (thickness $d=$ 500\,nm).
The chip contains several $Z-$shaped lines (``trapping wires'') of different widths used for trapping and moving the atoms, the broadest of which (\SI{100}{\micro \meter} wide) is used to trap atoms in the experiments described in this Letter.
The ring has an inner radius $r_\text{i} =$\SI{9}{\micro \meter} and an outer radius $r_\text{o} = $\SI{11}{\micro \meter}. 
It is placed \SI{70}{\micro \meter} from the edge of the trapping wire in $y$ direction.
The superconducting atom chip \cite{Bernon2013} is attached to the cold finger of a helium flow cryostat at temperature $T = $ \SI{4.2}{\kelvin}. 
The atoms are prepared in the hyperfine ground state 5S$_{1/2} F= 1, m_F = -1$ in a room temperature part of the setup and subsequently transported to a position below the superconducting chip by means of optical tweezers, as detailed in Ref.\ \cite{Cano2011}.

The microtrap is realized by the superposition of the fields generated by a current in the trapping wire and a homogeneous external bias field $\vec{B}_\text{bias}$.
An ensemble of $N\sim 1.5\times10^6$ atoms at $T_\text{atom}\sim$ \SI{1}{\micro \kelvin} is loaded from the optical tweezers into this superconducting microtrap, formed at \SI{400}{\micro \meter} from the chip surface.
After adiabatic compression, the cloud is evaporatively cooled to achieve either a thermal cloud or a nearly pure Bose-Einstein condensate (BEC).
The ensemble is then magnetically transported to a position $z \sim$ \SI{18}{\micro \meter} below the superconducting ring by rotating $\vec{B}_\text{bias}$ around the $x$ axis (Fig.\ref{fig:chip}(b)) and adjusting the current in the wire.
The longitudinal position of the cloud along $x$ is controlled by an additional field $\vec{B}_\text{conf}$ created by a confinement wire on the backside of the chip, see \cite{Bernon2013} for details.

The macroscopic superconducting ring exhibits quantum behavior that impacts the cold atomic cloud.
In the superconducting state the fluxoid is quantized \cite{London1948} as
\begin{equation}
n\cdot\Phi_0 =  \mu_0 \lambda_\text{L}^2\oint \vec{j}\cdot \text{d} \vec{s} + \Phi.
\end{equation}
This follows from the fact that the single-valuedness of the wavefunction requires any closed integral over the wave vector to be an integer multiple of $2\pi$.
Here, $\Phi_0 =h/2e$ is the magnetic flux quantum and the right hand side needs to be evaluated along a closed contour within the superconductor. 
$\lambda_\text{L}$ ($\sim$ \SI{100}{\nano\meter} for our Nb thin films) is the London pene\-tration depth, $\vec{j}$ is the supercurrent density and $\Phi$ is the total magnetic flux through the closed contour. 
If the superconductor is large compared to $\lambda_\text{L}$, which is the case for our geometry at temperatures well below the transition temperature $T_\text{c}$, the integral over $\vec{j}$ can be neglected. 
Then, $\Phi$ is quantized in multiples of $\Phi_0$:
\begin{equation}
n \cdot \Phi_0 =  \Phi = \int \vec{B}\cdot \text{d} \vec{A}.
\label{flux3}
\end{equation}
$\Phi$ is given by the sum of the flux applied above $T_\text{c}$, $\Phi_\text{freeze}=\int \vec{B}_\text{freeze} \cdot \text{d} \vec{A}$, and the  flux $LJ$ created by the supercurrents $J$ circulating around the ring 
\begin{equation}
n \cdot \Phi_0 =  \Phi_\text{freeze} + LJ.
\label{flux2}
\end{equation}
Here, $L$ is the inductance of the ring and $\vec{B}_\text{freeze}$ is the magnetic field applied to the ring during cooling. 
After cooling through $T_\text{c}$ the value of $n$ is defined as the integer closest to $\Phi_\text{freeze}/\Phi_0$.

Using $\Phi_0 = \Delta B_\text{freeze}\pi r_i r_o$ \cite{Brandt2004}, we expect for our geometry a field increment $\Delta B_\text{freeze}$ of about \SI{66.5}{mG} to change the flux in the ring by $1 \Phi_0$.
Having turned off $\vec{B}_\text{freeze}$, the (quantized) flux through the ring is conserved by the induced circulating current {$J_\text{freeze}$. 
Any fields applied to the ring in the superconducting state, for instance by the magnetic trap, are compensated by screening currents $J_\text{screen}$, so that the total current is $J=J_\text{freeze}+J_\text{screen}$.

\begin{figure}
\centerline{\includegraphics[width=.48\textwidth]{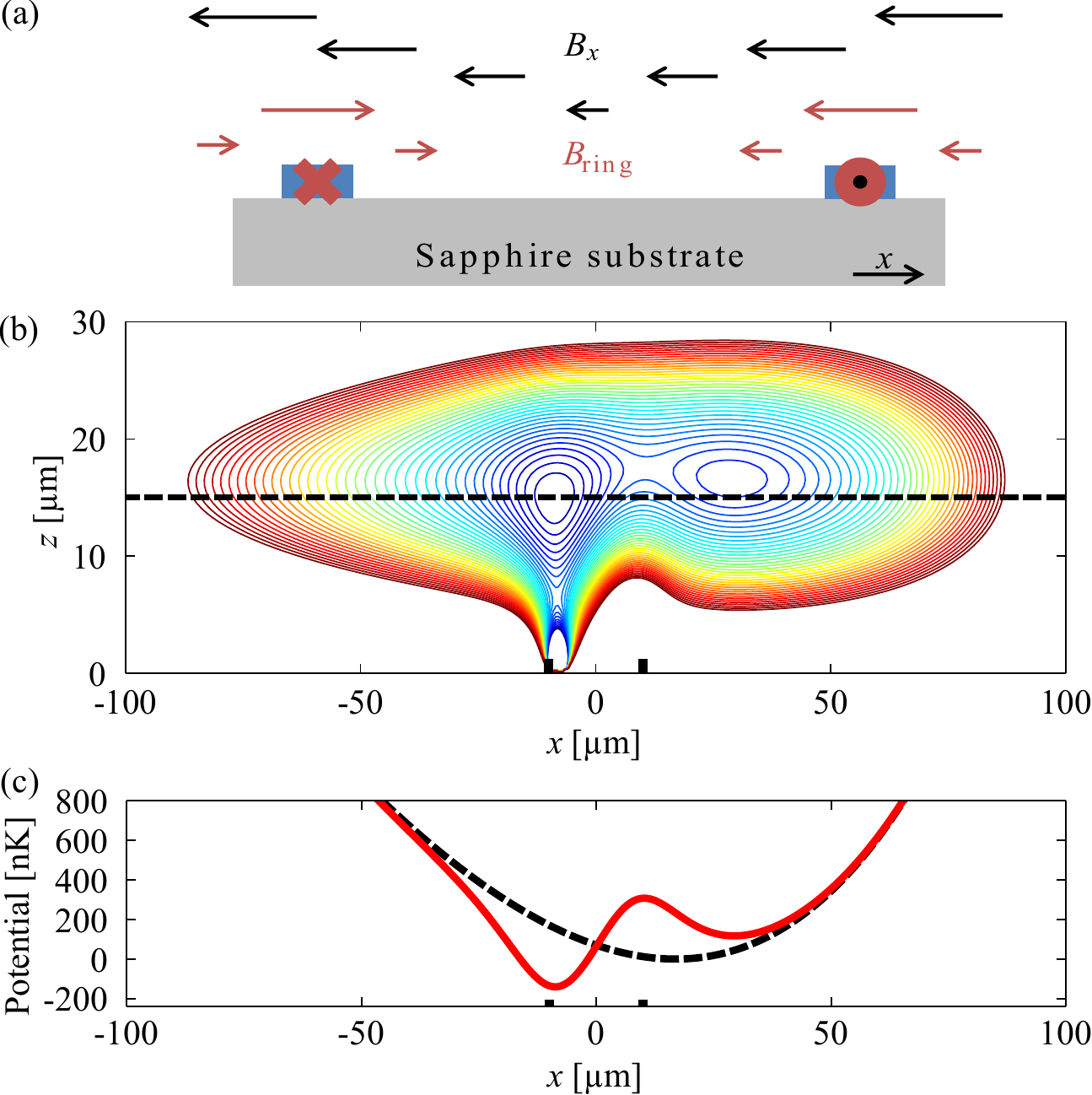}}
\caption{(a) Cross section along the dashed line in Fig. 1(a) with the principal magnetic $x$ field components of the trap and the ring in longitudinal direction. (b) Isopotential plot of the calculated trapping potential for 4 flux quanta in the ring.  An asymmetric potential with two local minima is created, with the lower minimum (dimple) above the ring structure (black markers on $x$ axis). Each contour line corresponds to an energy change of $k_\text{B} \cdot$\SI{50}{nK}. (c) Calculated potential along the longitudinal axis (black line Fig.\ \ref{fig:trap}(b)) with 4 flux quanta in the ring (solid red) and the unperturbed harmonic trap (dashed).}
\label{fig:trap}
\end{figure}
The magnetic (dipole) field $\vec{B}_\text{ring}$ created by currents $J$ locally modifies the magnetic trapping potential for the atoms in the vicinity of the structure.
To estimate the contribution of $\vec{B}_\text{ring}$ and its impact on the trapping potential, let us consider a cigar-shaped harmonic trap with oscillation frequencies $\omega_{x} \ll \omega_{y,z}$, whose radial axis $y$ is centered above the ring and whose size is on the order of the ring diameter.
The offset field $B_{x}$ at the minimum of the trap is considered to point along the $x$-axis.
The $x$-component of $\vec{B}_\text{ring}$ increases $B_{x}$ on one side of the ring and reduces it on the other (Fig.\  \ref{fig:trap}(a)), which leads to an asymmetric double well potential for the cold atomic cloud (Fig. \ref{fig:trap}(b) and (c)).
Hence, $\vec{B}_\text{ring}$ leads to a position shift of the minimum of the magnetic trapping potential along the longitudinal axis.
In addition, the position of the potential minimum is shifted  towards the surface with increasing number of flux quanta in the ring and the barrier height (trap depth) between the dimple and the surface is reduced.
This effect leads to a decrease in the number and temperature of atoms trappable in this dimple.
\begin{figure}
\centerline{\includegraphics[width=.48\textwidth]{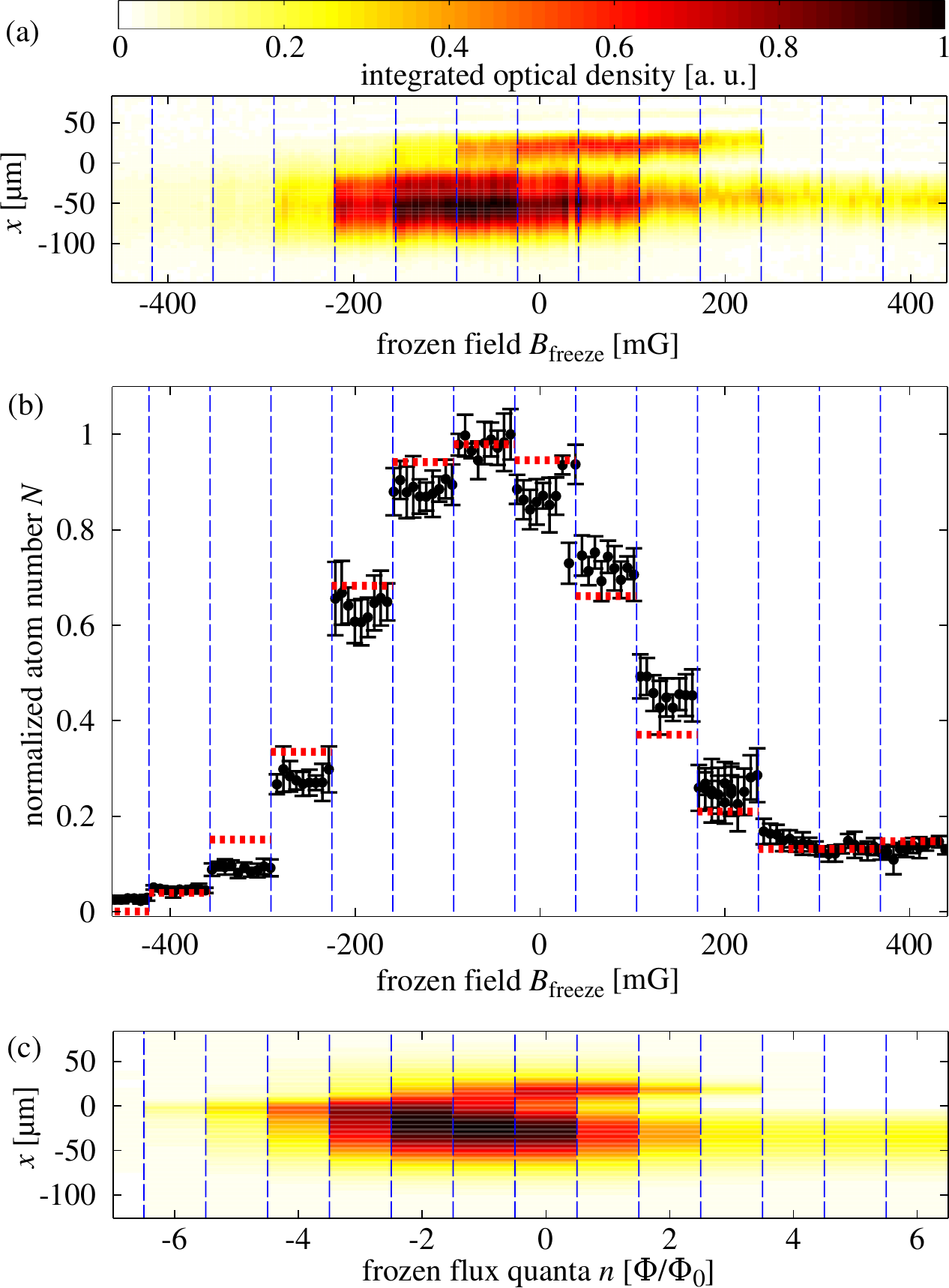}}
\caption{(a) Measured integrated density profile for different freezing fields. Each column represents the integrated density profile averaged over 9 absorption images. Adjacent lines differ by \SI{7}{\milli \gauss} in the field applied during cooling. The split in the density profile shows the emergence of the double well potential due to the ring field. (b) Relative atom number of an ensemble trapped at the superconducting ring, obtained by integrating the density profile along the $x$-axis shown in (a). The mean atom number is calculated from 9 pictures per frozen field applied during cooling. The dashed vertical lines have a spacing of $65.9 \,\si{\milli \gauss}$, which is the measured value for one flux quantum. The atom number is normalized to the maximum number measured in the trap. The red dashed lines indicate the calculated atom numbers obtained from the numerical simulation of the Boltzmann distribution.(c) Integrated density profile calculated from the predicted trapping potential using Boltzmann distributed atoms.}
\label{fig:density}
\end{figure}
Using the method described in \cite{Cano08} for the simulation of the supercurrent densities in the trapping wire and the ring, we numerically calculate the field distribution composed of the ring field and the trapping field with Biot-Savart's law.
The results of these calculations are in good agreement with simulations based on inductance calculations (3D-MLSI software package) \cite{khap}.
Below the ring structure, the ring field for 1$\Phi_0$ leads to a field shift of $\sim$ \SI{3}{mG} as compared with the unperturbed harmonic trap (Fig.\ \ref{fig:trap}(c)), corresponding to a dimple with a depth on the order of \SI{100}{\nano \kelvin}.
Furthermore, the alteration of the potential landscape caused by the circular supercurrents leads to a longitudinal center-of-mass oscillation frequency  that depends on the number of flux quanta $n$ in the ring.

For the measurements, we first prepare the flux state of the ring by heating up the chip to a temperature above $T_\text{c}$ and subsequently cool it to $T=$ \SI{4.2}{\kelvin} in a homogeneous magnetic field $\vec{B}_\text{freeze}$,  applied perpendicular to the surface and ranging from \SI{-500}{mG} to \SI{500}{mG}.
The magnetic fields are calibrated by microwave spectroscopy, i.e., the atoms are prepared in the state $F = 1, m_F = -1$ and the number of atoms in the state $F = 2, m_F = 0$ is measured after application of a microwave pulse with variable frequency.
Limited by fluctuations of the magnetic field in the laboratory, the absolute value of $B_\text{freeze}$ is known within $\pm$\SI{5}{\milli \gauss}.

A thermal cloud of $N \sim 2\times10^5$ atoms with a temperature of $\sim$ \SI{250}{\nano \kelvin} is prepared below the trapping wire and brought close to the ring, where it is held for \SI{1}{s}.
For each value of $B_\text{freeze}$ we take nine absorption images \textit{in situ} by reflection imaging \cite{Smith2011} along the $y$ direction. 
After averaging over the images, we integrate the calculated column density along the $z$ direction to obtain a one dimensional profile of the atomic cloud along the axis of weak confinement to reveal the impact of the ring field along $x$.
In Fig.\ \ref{fig:density}(a) the density profiles are plotted vs.\ $B_\text{freeze}$.
We observe steps in the integrated density profile occurring when the number of flux quanta in the ring changes.
For certain flux states, two  distinct density peaks, which indicate the double well potential, are discernible.
By further integration of the profiles shown in Fig.\ \ref{fig:density}(a), we obtain the atom number $N$ as a function of $B_\text{freeze}$.
The atom number is normalized to the maximum number measured in the trap and plotted in Fig.\ \ref{fig:density}(b). 
There are clearly visible equidistant steps with a width of $\Delta B_\text{freeze} = 65.9 \pm 2.3\,\si{\milli \gauss}$, indicated by the blue vertical lines. 
The theoretically predicted value of $\Delta B_\text{freeze} = \SI{66.5}{\milli \gauss}$ per flux quantum is well within the error bars of the measurement.
In Fig.\ \ref{fig:density}(b) it is visible that we achieve a resolution better than single flux quanta.

The measurements in Fig.\ \ref{fig:density}(b) are not symmetric around the value $B_\text{freeze} =0$.
As the magnetic trap itself has a magnetic field component perpendicular to the surface, a screening current $J_\text{screen}$ is induced in the ring to compensate this field.
The screening current contributes to the trapping potential even for $B_\text{freeze} =0$.  
Only if the sum of the fields perpendicular to the ring is equal to the number of trapped flux quanta, the harmonic trap is unperturbed by the screening currents.
In this case, there is no net current around the ring, i.e.\ $J_\text{screen} = -J_\text{freeze}$, and only Meissner currents, which keep the superconducting film itself field free, are present \cite{Cano08}.

To gain a qualitative understanding of the impact of the screening currents on the density profile, we set up a simplified model and numerically calculate the modification of the trapping potential for different numbers of flux quanta in the ring.
We simulate the atomic density in the trap using a Maxwell-Boltzmann distribution for the energy, assuming a maximum temperature of $T_\text{max} = 230$\,nK.
As the barrier height between the trap and the surface (trap depth) depends on the number of flux quanta $n$ in the ring, we truncate the energy in the Boltzmann distribution to the trap depth, which changes the volume occupied by the cloud \cite{Ketterle96, Davis95b, Maerkle2014}.
Furthermore, we incorporate the loss of atoms according to a heuristic scaling of the atomic density, assuming $\rho(n)/\rho_\text{max}=T_\text{depth}(n)/T_\text{depth}(0)$.
The simulations take into account the existence of a double well potential, in which both traps have different trap depths.
To compare the calculations with the observed density profiles, we sum over the calculated atomic density along the $y$ and $z$ directions and plot the result vs. the number of frozen flux quanta $n$.
The result is shown in Fig.\ \ref{fig:density}(c), which closely resembles the experimental data in Fig.\ \ref{fig:density}(a).
The number of trapped atoms is estimated by additionally summing up the density distribution (Fig.\ \ref{fig:density}(c)) along the $x$-direction, which leads to the red dashed lines in Fig.\ \ref{fig:density}(b).
Our simple model qualitatively and moderately quantitatively matches the behavior of the measured atom number for different flux quanta $n$ and justifies the assumed scaling in the densities.
We attribute the discrepancies between experiment and simulation to losses during the loading process of the double well, which are not taken into account in the calculations. 
\begin{figure}
\centerline{\includegraphics[width=.5\textwidth]{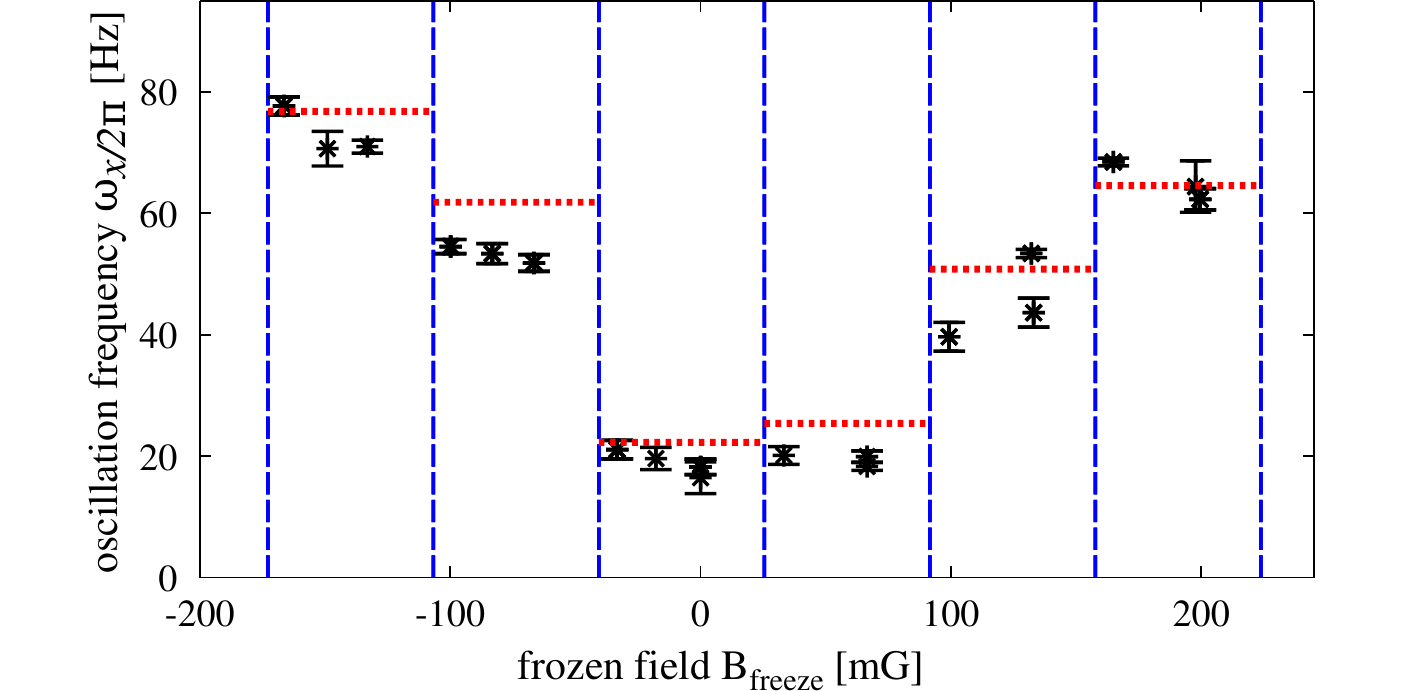}}
\caption{Trapping frequencies measured for different freezing fields. The dashed vertical lines indicate the jump in the flux quanta. The black dots with errorbars were obtained from the measurement. The dotted horizontal lines are calculated values for different numbers of frozen flux quanta.}
\label{fig:freq}
\end{figure}

In order to obtain additional information on the number of frozen flux quanta, we have performed a measurement of the center-of-mass oscillation frequency of the trap at the ring for various values of $B_\text{freeze}$.
For this measurement, we prepared a BEC filling only the low-lying dimple of the potential with atoms, where the trap frequency is expected to vary significantly with the number of flux quanta.
To measure the frequency, a center-of-mass oscillation of the atoms along the longitudinal axis of the trap was excited by rapidly displacing the minimum of the magnetic potential using the current in the confinement wire.
After a variable hold time (0 to 200\,ms), a microwave pulse was applied to transfer the atoms into the untrapped $F = 2, m_F = 0$ state and the oscillation frequency was extracted from the position of the cloud after a time-of-flight of \SI{12}{ms}.
Fig.\ \ref{fig:freq} shows the measured oscillation frequencies along the expected values extracted from our simulation of the potential.
The dotted vertical lines are based on an atom number  measurement similar as in Fig.\ \ref{fig:density}(b) and show the expected values of $B_\text{freeze}$ at which the number of flux quanta in the ring changes.
The simulations are in qualitative agreement with the measurement and show our resolution on the level of single flux quanta; deviations between experiment and calculation can be attributed to uncertainties in the applied magnetic fields.

In summary, we have demonstrated that a cold atomic cloud of $^{87}$Rb atoms positioned close to a superconducting ring is sensitive to the magnetic field created by single flux quanta.
The modification of the trapping potential by this field is detectable in two trap characteristics, firstly in the trap depth and therefore in the atom number of the ensemble, and secondly in the trapping frequency inside the created dimple trap.
We also expect that the variation of the number of flux quanta in the ring will impact internal degrees of freedom of the trapped atoms, such as the energy difference between Zeeman sublevels, which is accessible by means of Ramsey interferometry.
This sensitivity paves the way towards future experiments that interface, e.g., SQUIDs and cold atomic clouds and exploit the atomic ensemble as a robust quantum memory \cite{Patton2013b}.
Strong coupling, as demonstrated for example with Nitrogen vacancy centers \cite{Zhu2011}, could be achieved for cold atoms by reducing the loop size, resulting in larger fields per flux quantum, or by bosonic enhancement of the coupling strength between two macroscopically populated atomic states \cite{Patton2013a}.
The impact of the applied magnetic fields, trapped flux within the superconducting structures, and stray light on the coherence of the superconducting circuits is yet to be studied, but is not expected to fundamentally limit the coupling between the two systems.
\begin{acknowledgments}
The authors would like to thank Thomas Dahm and Igor Sapina for useful discussions.
This work was supported the Deutsche Forschungsgemeinschaft (SFB TRR21) and by the European Research Council (Socathes). 
The authors acknowledge additional support from the ev.\ Studienwerk Villigst e.V., and the Carl Zeiss Stiftung. 
\end{acknowledgments}


\begin{thebibliography}{44}%
\makeatletter
\providecommand \@ifxundefined [1]{%
 \@ifx{#1\undefined}
}%
\providecommand \@ifnum [1]{%
 \ifnum #1\expandafter \@firstoftwo
 \else \expandafter \@secondoftwo
 \fi
}%
\providecommand \@ifx [1]{%
 \ifx #1\expandafter \@firstoftwo
 \else \expandafter \@secondoftwo
 \fi
}%
\providecommand \natexlab [1]{#1}%
\providecommand \enquote  [1]{``#1''}%
\providecommand \bibnamefont  [1]{#1}%
\providecommand \bibfnamefont [1]{#1}%
\providecommand \citenamefont [1]{#1}%
\providecommand \href@noop [0]{\@secondoftwo}%
\providecommand \href [0]{\begingroup \@sanitize@url \@href}%
\providecommand \@href[1]{\@@startlink{#1}\@@href}%
\providecommand \@@href[1]{\endgroup#1\@@endlink}%
\providecommand \@sanitize@url [0]{\catcode `\\12\catcode `\$12\catcode
  `\&12\catcode `\#12\catcode `\^12\catcode `\_12\catcode `\%12\relax}%
\providecommand \@@startlink[1]{}%
\providecommand \@@endlink[0]{}%
\providecommand \url  [0]{\begingroup\@sanitize@url \@url }%
\providecommand \@url [1]{\endgroup\@href {#1}{\urlprefix }}%
\providecommand \urlprefix  [0]{URL }%
\providecommand \Eprint [0]{\href }%
\providecommand \doibase [0]{http://dx.doi.org/}%
\providecommand \selectlanguage [0]{\@gobble}%
\providecommand \bibinfo  [0]{\@secondoftwo}%
\providecommand \bibfield  [0]{\@secondoftwo}%
\providecommand \translation [1]{[#1]}%
\providecommand \BibitemOpen [0]{}%
\providecommand \bibitemStop [0]{}%
\providecommand \bibitemNoStop [0]{.\EOS\space}%
\providecommand \EOS [0]{\spacefactor3000\relax}%
\providecommand \BibitemShut  [1]{\csname bibitem#1\endcsname}%
\let\auto@bib@innerbib\@empty
\bibitem [{\citenamefont {Rabl}\ \emph {et~al.}(2006)\citenamefont {Rabl},
  \citenamefont {DeMille}, \citenamefont {Doyle}, \citenamefont {Lukin},
  \citenamefont {Schoelkopf},\ and\ \citenamefont {Zoller}}]{Rabl2006}%
  \BibitemOpen
  \bibfield  {author} {\bibinfo {author} {\bibfnamefont {P.}~\bibnamefont
  {Rabl}}, \bibinfo {author} {\bibfnamefont {D.}~\bibnamefont {DeMille}},
  \bibinfo {author} {\bibfnamefont {J.~M.}\ \bibnamefont {Doyle}}, \bibinfo
  {author} {\bibfnamefont {M.~D.}\ \bibnamefont {Lukin}}, \bibinfo {author}
  {\bibfnamefont {R.~J.}\ \bibnamefont {Schoelkopf}}, \ and\ \bibinfo {author}
  {\bibfnamefont {P.}~\bibnamefont {Zoller}},\ }\href {\doibase
  10.1103/PhysRevLett.97.033003} {\bibfield  {journal} {\bibinfo  {journal}
  {Phys. Rev. Lett.}\ }\textbf {\bibinfo {volume} {97}},\ \bibinfo {pages}
  {033003} (\bibinfo {year} {2006})}\BibitemShut {NoStop}%
\bibitem [{\citenamefont {Petrosyan}\ and\ \citenamefont
  {Fleischhauer}(2008)}]{Petrosyan2008}%
  \BibitemOpen
  \bibfield  {author} {\bibinfo {author} {\bibfnamefont {D.}~\bibnamefont
  {Petrosyan}}\ and\ \bibinfo {author} {\bibfnamefont {M.}~\bibnamefont
  {Fleischhauer}},\ }\href {\doibase 10.1103/PhysRevLett.100.170501} {\bibfield
   {journal} {\bibinfo  {journal} {Phys. Rev. Lett.}\ }\textbf {\bibinfo
  {volume} {100}},\ \bibinfo {pages} {170501} (\bibinfo {year}
  {2008})}\BibitemShut {NoStop}%
\bibitem [{\citenamefont {Petrosyan}\ \emph {et~al.}(2009)\citenamefont
  {Petrosyan}, \citenamefont {Bensky}, \citenamefont {Kurizki}, \citenamefont
  {Mazets}, \citenamefont {Majer},\ and\ \citenamefont
  {Schmiedmayer}}]{Petrosyan2009}%
  \BibitemOpen
  \bibfield  {author} {\bibinfo {author} {\bibfnamefont {D.}~\bibnamefont
  {Petrosyan}}, \bibinfo {author} {\bibfnamefont {G.}~\bibnamefont {Bensky}},
  \bibinfo {author} {\bibfnamefont {G.}~\bibnamefont {Kurizki}}, \bibinfo
  {author} {\bibfnamefont {I.}~\bibnamefont {Mazets}}, \bibinfo {author}
  {\bibfnamefont {J.}~\bibnamefont {Majer}}, \ and\ \bibinfo {author}
  {\bibfnamefont {J.}~\bibnamefont {Schmiedmayer}},\ }\href {\doibase
  10.1103/PhysRevA.79.040304} {\bibfield  {journal} {\bibinfo  {journal} {Phys.
  Rev. A}\ }\textbf {\bibinfo {volume} {79}},\ \bibinfo {pages} {040304}
  (\bibinfo {year} {2009})}\BibitemShut {NoStop}%
\bibitem [{\citenamefont {Verd\'u}\ \emph {et~al.}(2009)\citenamefont
  {Verd\'u}, \citenamefont {Zoubi}, \citenamefont {Koller}, \citenamefont
  {Majer}, \citenamefont {Ritsch},\ and\ \citenamefont
  {Schmiedmayer}}]{Verdu2009}%
  \BibitemOpen
  \bibfield  {author} {\bibinfo {author} {\bibfnamefont {J.}~\bibnamefont
  {Verd\'u}}, \bibinfo {author} {\bibfnamefont {H.}~\bibnamefont {Zoubi}},
  \bibinfo {author} {\bibfnamefont {C.}~\bibnamefont {Koller}}, \bibinfo
  {author} {\bibfnamefont {J.}~\bibnamefont {Majer}}, \bibinfo {author}
  {\bibfnamefont {H.}~\bibnamefont {Ritsch}}, \ and\ \bibinfo {author}
  {\bibfnamefont {J.}~\bibnamefont {Schmiedmayer}},\ }\href {\doibase
  10.1103/PhysRevLett.103.043603} {\bibfield  {journal} {\bibinfo  {journal}
  {Phys. Rev. Lett.}\ }\textbf {\bibinfo {volume} {103}},\ \bibinfo {pages}
  {043603} (\bibinfo {year} {2009})}\BibitemShut {NoStop}%
\bibitem [{\citenamefont {Bernon}\ \emph {et~al.}(2013)\citenamefont {Bernon},
  \citenamefont {Hattermann}, \citenamefont {Bothner}, \citenamefont
  {Knufinke}, \citenamefont {Weiss}, \citenamefont {Jessen}, \citenamefont
  {Cano}, \citenamefont {Kemmler}, \citenamefont {Kleiner}, \citenamefont
  {Koelle},\ and\ \citenamefont {Fort\'agh}}]{Bernon2013}%
  \BibitemOpen
  \bibfield  {author} {\bibinfo {author} {\bibfnamefont {S.}~\bibnamefont
  {Bernon}}, \bibinfo {author} {\bibfnamefont {H.}~\bibnamefont {Hattermann}},
  \bibinfo {author} {\bibfnamefont {D.}~\bibnamefont {Bothner}}, \bibinfo
  {author} {\bibfnamefont {M.}~\bibnamefont {Knufinke}}, \bibinfo {author}
  {\bibfnamefont {P.}~\bibnamefont {Weiss}}, \bibinfo {author} {\bibfnamefont
  {F.}~\bibnamefont {Jessen}}, \bibinfo {author} {\bibfnamefont
  {D.}~\bibnamefont {Cano}}, \bibinfo {author} {\bibfnamefont {M.}~\bibnamefont
  {Kemmler}}, \bibinfo {author} {\bibfnamefont {R.}~\bibnamefont {Kleiner}},
  \bibinfo {author} {\bibfnamefont {D.}~\bibnamefont {Koelle}}, \ and\ \bibinfo
  {author} {\bibfnamefont {J.}~\bibnamefont {Fort\'agh}},\ }\href {\doibase
  10.1038/ncomms3380} {\bibfield  {journal} {\bibinfo  {journal} {Nat.
  Commun.}\ }\textbf {\bibinfo {volume} {4}},\ \bibinfo {eid} {2380} (\bibinfo
  {year} {2013}),\ 10.1038/ncomms3380}\BibitemShut {NoStop}%
\bibitem [{\citenamefont {Patton}\ and\ \citenamefont
  {Fischer}(2013{\natexlab{a}})}]{Patton2013}%
  \BibitemOpen
  \bibfield  {author} {\bibinfo {author} {\bibfnamefont {K.~R.}\ \bibnamefont
  {Patton}}\ and\ \bibinfo {author} {\bibfnamefont {U.~R.}\ \bibnamefont
  {Fischer}},\ }\href {\doibase 10.1103/PhysRevA.87.052303} {\bibfield
  {journal} {\bibinfo  {journal} {Phys. Rev. A}\ }\textbf {\bibinfo {volume}
  {87}},\ \bibinfo {pages} {052303} (\bibinfo {year}
  {2013}{\natexlab{a}})}\BibitemShut {NoStop}%
\bibitem [{\citenamefont {Hogan}\ \emph {et~al.}(2012)\citenamefont {Hogan},
  \citenamefont {Agner}, \citenamefont {Merkt}, \citenamefont {Thiele},
  \citenamefont {Filipp},\ and\ \citenamefont {Wallraff}}]{Hogan2012}%
  \BibitemOpen
  \bibfield  {author} {\bibinfo {author} {\bibfnamefont {S.~D.}\ \bibnamefont
  {Hogan}}, \bibinfo {author} {\bibfnamefont {J.~A.}\ \bibnamefont {Agner}},
  \bibinfo {author} {\bibfnamefont {F.}~\bibnamefont {Merkt}}, \bibinfo
  {author} {\bibfnamefont {T.}~\bibnamefont {Thiele}}, \bibinfo {author}
  {\bibfnamefont {S.}~\bibnamefont {Filipp}}, \ and\ \bibinfo {author}
  {\bibfnamefont {A.}~\bibnamefont {Wallraff}},\ }\href {\doibase
  10.1103/PhysRevLett.108.063004} {\bibfield  {journal} {\bibinfo  {journal}
  {Phys. Rev. Lett.}\ }\textbf {\bibinfo {volume} {108}},\ \bibinfo {pages}
  {063004} (\bibinfo {year} {2012})}\BibitemShut {NoStop}%
\bibitem [{\citenamefont {Kim}\ \emph {et~al.}(2011)\citenamefont {Kim},
  \citenamefont {Vlahacos}, \citenamefont {Hoffman}, \citenamefont {Grover},
  \citenamefont {Voigt}, \citenamefont {Cooper}, \citenamefont {Ballard},
  \citenamefont {Palmer}, \citenamefont {Hafezi}, \citenamefont {Taylor},
  \citenamefont {Anderson}, \citenamefont {Dragt}, \citenamefont {Lobb},
  \citenamefont {Orozco}, \citenamefont {Rolston},\ and\ \citenamefont
  {Wellstood}}]{Kim2011Wellstood}%
  \BibitemOpen
  \bibfield  {author} {\bibinfo {author} {\bibfnamefont {Z.}~\bibnamefont
  {Kim}}, \bibinfo {author} {\bibfnamefont {C.~P.}\ \bibnamefont {Vlahacos}},
  \bibinfo {author} {\bibfnamefont {J.~E.}\ \bibnamefont {Hoffman}}, \bibinfo
  {author} {\bibfnamefont {J.~A.}\ \bibnamefont {Grover}}, \bibinfo {author}
  {\bibfnamefont {K.~D.}\ \bibnamefont {Voigt}}, \bibinfo {author}
  {\bibfnamefont {B.~K.}\ \bibnamefont {Cooper}}, \bibinfo {author}
  {\bibfnamefont {C.~J.}\ \bibnamefont {Ballard}}, \bibinfo {author}
  {\bibfnamefont {B.~S.}\ \bibnamefont {Palmer}}, \bibinfo {author}
  {\bibfnamefont {M.}~\bibnamefont {Hafezi}}, \bibinfo {author} {\bibfnamefont
  {J.~M.}\ \bibnamefont {Taylor}}, \bibinfo {author} {\bibfnamefont {J.~R.}\
  \bibnamefont {Anderson}}, \bibinfo {author} {\bibfnamefont {A.~J.}\
  \bibnamefont {Dragt}}, \bibinfo {author} {\bibfnamefont {C.~J.}\ \bibnamefont
  {Lobb}}, \bibinfo {author} {\bibfnamefont {L.~A.}\ \bibnamefont {Orozco}},
  \bibinfo {author} {\bibfnamefont {S.~L.}\ \bibnamefont {Rolston}}, \ and\
  \bibinfo {author} {\bibfnamefont {F.~C.}\ \bibnamefont {Wellstood}},\ }\href
  {\doibase http://dx.doi.org/10.1063/1.3651466} {\bibfield  {journal}
  {\bibinfo  {journal} {AIP Adv.}\ }\textbf {\bibinfo {volume} {1}},\ \bibinfo
  {eid} {042107} (\bibinfo {year} {2011})}\BibitemShut {NoStop}%
\bibitem [{\citenamefont {Wallquist}\ \emph {et~al.}(2009)\citenamefont
  {Wallquist}, \citenamefont {Hammerer}, \citenamefont {Rabl}, \citenamefont
  {Lukin},\ and\ \citenamefont {Zoller}}]{Wallquist2009}%
  \BibitemOpen
  \bibfield  {author} {\bibinfo {author} {\bibfnamefont {M.}~\bibnamefont
  {Wallquist}}, \bibinfo {author} {\bibfnamefont {K.}~\bibnamefont {Hammerer}},
  \bibinfo {author} {\bibfnamefont {P.}~\bibnamefont {Rabl}}, \bibinfo {author}
  {\bibfnamefont {M.}~\bibnamefont {Lukin}}, \ and\ \bibinfo {author}
  {\bibfnamefont {P.}~\bibnamefont {Zoller}},\ }\href
  {http://stacks.iop.org/1402-4896/2009/i=T137/a=014001} {\bibfield  {journal}
  {\bibinfo  {journal} {Phys. Scripta}\ }\textbf {\bibinfo {volume} {2009}},\
  \bibinfo {pages} {014001} (\bibinfo {year} {2009})}\BibitemShut {NoStop}%
\bibitem [{\citenamefont {Hafezi}\ \emph {et~al.}(2012)\citenamefont {Hafezi},
  \citenamefont {Kim}, \citenamefont {Rolston}, \citenamefont {Orozco},
  \citenamefont {Lev},\ and\ \citenamefont {Taylor}}]{Hafezi2012}%
  \BibitemOpen
  \bibfield  {author} {\bibinfo {author} {\bibfnamefont {M.}~\bibnamefont
  {Hafezi}}, \bibinfo {author} {\bibfnamefont {Z.}~\bibnamefont {Kim}},
  \bibinfo {author} {\bibfnamefont {S.~L.}\ \bibnamefont {Rolston}}, \bibinfo
  {author} {\bibfnamefont {L.~A.}\ \bibnamefont {Orozco}}, \bibinfo {author}
  {\bibfnamefont {B.~L.}\ \bibnamefont {Lev}}, \ and\ \bibinfo {author}
  {\bibfnamefont {J.~M.}\ \bibnamefont {Taylor}},\ }\href {\doibase
  10.1103/PhysRevA.85.020302} {\bibfield  {journal} {\bibinfo  {journal} {Phys.
  Rev. A}\ }\textbf {\bibinfo {volume} {85}},\ \bibinfo {pages} {020302}
  (\bibinfo {year} {2012})}\BibitemShut {NoStop}%
\bibitem [{\citenamefont {Jessen}\ \emph {et~al.}(2014)\citenamefont {Jessen},
  \citenamefont {Knufinke}, \citenamefont {Bell}, \citenamefont {Vergien},
  \citenamefont {Hattermann}, \citenamefont {Weiss}, \citenamefont {Rudolph},
  \citenamefont {Reinschmidt}, \citenamefont {Meyer}, \citenamefont {Gaber},
  \citenamefont {Cano}, \citenamefont {Günther}, \citenamefont {Bernon},
  \citenamefont {Koelle}, \citenamefont {Kleiner},\ and\ \citenamefont
  {Fortágh}}]{jessen2013}%
  \BibitemOpen
  \bibfield  {author} {\bibinfo {author} {\bibfnamefont {F.}~\bibnamefont
  {Jessen}}, \bibinfo {author} {\bibfnamefont {M.}~\bibnamefont {Knufinke}},
  \bibinfo {author} {\bibfnamefont {S.}~\bibnamefont {Bell}}, \bibinfo {author}
  {\bibfnamefont {P.}~\bibnamefont {Vergien}}, \bibinfo {author} {\bibfnamefont
  {H.}~\bibnamefont {Hattermann}}, \bibinfo {author} {\bibfnamefont
  {P.}~\bibnamefont {Weiss}}, \bibinfo {author} {\bibfnamefont
  {M.}~\bibnamefont {Rudolph}}, \bibinfo {author} {\bibfnamefont
  {M.}~\bibnamefont {Reinschmidt}}, \bibinfo {author} {\bibfnamefont
  {K.}~\bibnamefont {Meyer}}, \bibinfo {author} {\bibfnamefont
  {T.}~\bibnamefont {Gaber}}, \bibinfo {author} {\bibfnamefont
  {D.}~\bibnamefont {Cano}}, \bibinfo {author} {\bibfnamefont {A.}~\bibnamefont
  {Günther}}, \bibinfo {author} {\bibfnamefont {S.}~\bibnamefont {Bernon}},
  \bibinfo {author} {\bibfnamefont {D.}~\bibnamefont {Koelle}}, \bibinfo
  {author} {\bibfnamefont {R.}~\bibnamefont {Kleiner}}, \ and\ \bibinfo
  {author} {\bibfnamefont {J.}~\bibnamefont {Fortágh}},\ }\href {\doibase
  10.1007/s00340-013-5750-5} {\bibfield  {journal} {\bibinfo  {journal} {Appl.
  Phys. B}\ }\textbf {\bibinfo {volume} {116}},\ \bibinfo {pages} {665}
  (\bibinfo {year} {2014})}\BibitemShut {NoStop}%
\bibitem [{\citenamefont {Minniberger}\ \emph {et~al.}(2014)\citenamefont
  {Minniberger}, \citenamefont {Diorico}, \citenamefont {Haslinger},
  \citenamefont {Hufnagel}, \citenamefont {Novotny}, \citenamefont {Lippok},
  \citenamefont {Majer}, \citenamefont {Koller}, \citenamefont {Schneider},\
  and\ \citenamefont {Schmiedmayer}}]{Minniberger2014}%
  \BibitemOpen
  \bibfield  {author} {\bibinfo {author} {\bibfnamefont {S.}~\bibnamefont
  {Minniberger}}, \bibinfo {author} {\bibfnamefont {F.}~\bibnamefont
  {Diorico}}, \bibinfo {author} {\bibfnamefont {S.}~\bibnamefont {Haslinger}},
  \bibinfo {author} {\bibfnamefont {C.}~\bibnamefont {Hufnagel}}, \bibinfo
  {author} {\bibfnamefont {C.}~\bibnamefont {Novotny}}, \bibinfo {author}
  {\bibfnamefont {N.}~\bibnamefont {Lippok}}, \bibinfo {author} {\bibfnamefont
  {J.}~\bibnamefont {Majer}}, \bibinfo {author} {\bibfnamefont
  {C.}~\bibnamefont {Koller}}, \bibinfo {author} {\bibfnamefont
  {S.}~\bibnamefont {Schneider}}, \ and\ \bibinfo {author} {\bibfnamefont
  {J.}~\bibnamefont {Schmiedmayer}},\ }\href {\doibase
  10.1007/s00340-014-5790-5} {\bibfield  {journal} {\bibinfo  {journal} {Appl.
  Phys. B}\ }\textbf {\bibinfo {volume} {116}},\ \bibinfo {pages} {1017}
  (\bibinfo {year} {2014})}\BibitemShut {NoStop}%
\bibitem [{\citenamefont {Pritchard}\ \emph {et~al.}(2014)\citenamefont
  {Pritchard}, \citenamefont {Isaacs}, \citenamefont {Beck}, \citenamefont
  {McDermott},\ and\ \citenamefont {Saffman}}]{Pritchard2014}%
  \BibitemOpen
  \bibfield  {author} {\bibinfo {author} {\bibfnamefont {J.~D.}\ \bibnamefont
  {Pritchard}}, \bibinfo {author} {\bibfnamefont {J.~A.}\ \bibnamefont
  {Isaacs}}, \bibinfo {author} {\bibfnamefont {M.~A.}\ \bibnamefont {Beck}},
  \bibinfo {author} {\bibfnamefont {R.}~\bibnamefont {McDermott}}, \ and\
  \bibinfo {author} {\bibfnamefont {M.}~\bibnamefont {Saffman}},\ }\href
  {\doibase 10.1103/PhysRevA.89.010301} {\bibfield  {journal} {\bibinfo
  {journal} {Phys. Rev. A}\ }\textbf {\bibinfo {volume} {89}},\ \bibinfo
  {pages} {010301} (\bibinfo {year} {2014})}\BibitemShut {NoStop}%
\bibitem [{\citenamefont {Scheel}\ \emph {et~al.}(2005)\citenamefont {Scheel},
  \citenamefont {Rekdal}, \citenamefont {Knight},\ and\ \citenamefont
  {Hinds}}]{Scheel05}%
  \BibitemOpen
  \bibfield  {author} {\bibinfo {author} {\bibfnamefont {S.}~\bibnamefont
  {Scheel}}, \bibinfo {author} {\bibfnamefont {P.~K.}\ \bibnamefont {Rekdal}},
  \bibinfo {author} {\bibfnamefont {P.~L.}\ \bibnamefont {Knight}}, \ and\
  \bibinfo {author} {\bibfnamefont {E.~A.}\ \bibnamefont {Hinds}},\ }\href
  {\doibase 10.1103/PhysRevA.72.042901} {\bibfield  {journal} {\bibinfo
  {journal} {Phys. Rev. A}\ }\textbf {\bibinfo {volume} {72}},\ \bibinfo {eid}
  {042901} (\bibinfo {year} {2005})}\BibitemShut {NoStop}%
\bibitem [{\citenamefont {Skagerstam}\ \emph {et~al.}(2006)\citenamefont
  {Skagerstam}, \citenamefont {U.Hohenester}, \citenamefont {Eiguren},\ and\
  \citenamefont {Rekdal}}]{Skagerstam06}%
  \BibitemOpen
  \bibfield  {author} {\bibinfo {author} {\bibfnamefont {B.~K.}\ \bibnamefont
  {Skagerstam}}, \bibinfo {author} {\bibnamefont {U.Hohenester}}, \bibinfo
  {author} {\bibfnamefont {A.}~\bibnamefont {Eiguren}}, \ and\ \bibinfo
  {author} {\bibfnamefont {P.~K.}\ \bibnamefont {Rekdal}},\ }\href {\doibase
  10.1103/PhysRevLett.97.070401} {\bibfield  {journal} {\bibinfo  {journal}
  {Phys. Rev. Lett}\ }\textbf {\bibinfo {volume} {97}},\ \bibinfo {eid}
  {070401} (\bibinfo {year} {2006})}\BibitemShut {NoStop}%
\bibitem [{\citenamefont {Cano}\ \emph
  {et~al.}(2008{\natexlab{a}})\citenamefont {Cano}, \citenamefont {Kasch},
  \citenamefont {Hattermann}, \citenamefont {Kleiner}, \citenamefont
  {Zimmermann}, \citenamefont {Koelle},\ and\ \citenamefont
  {Fort\'agh}}]{Cano08a}%
  \BibitemOpen
  \bibfield  {author} {\bibinfo {author} {\bibfnamefont {D.}~\bibnamefont
  {Cano}}, \bibinfo {author} {\bibfnamefont {B.}~\bibnamefont {Kasch}},
  \bibinfo {author} {\bibfnamefont {H.}~\bibnamefont {Hattermann}}, \bibinfo
  {author} {\bibfnamefont {R.}~\bibnamefont {Kleiner}}, \bibinfo {author}
  {\bibfnamefont {C.}~\bibnamefont {Zimmermann}}, \bibinfo {author}
  {\bibfnamefont {D.}~\bibnamefont {Koelle}}, \ and\ \bibinfo {author}
  {\bibfnamefont {J.}~\bibnamefont {Fort\'agh}},\ }\href {\doibase
  10.1103/PhysRevLett.101.183006} {\bibfield  {journal} {\bibinfo  {journal}
  {Phys. Rev. Lett.}\ }\textbf {\bibinfo {volume} {101}},\ \bibinfo {pages}
  {183006} (\bibinfo {year} {2008}{\natexlab{a}})}\BibitemShut {NoStop}%
\bibitem [{\citenamefont {Kasch}\ \emph {et~al.}(2010)\citenamefont {Kasch},
  \citenamefont {Hattermann}, \citenamefont {Cano}, \citenamefont {Judd},
  \citenamefont {Scheel}, \citenamefont {Zimmermann}, \citenamefont {Kleiner},
  \citenamefont {Koelle},\ and\ \citenamefont {Fort\'agh}}]{Kasch2010}%
  \BibitemOpen
  \bibfield  {author} {\bibinfo {author} {\bibfnamefont {B.}~\bibnamefont
  {Kasch}}, \bibinfo {author} {\bibfnamefont {H.}~\bibnamefont {Hattermann}},
  \bibinfo {author} {\bibfnamefont {D.}~\bibnamefont {Cano}}, \bibinfo {author}
  {\bibfnamefont {T.~E.}\ \bibnamefont {Judd}}, \bibinfo {author}
  {\bibfnamefont {S.}~\bibnamefont {Scheel}}, \bibinfo {author} {\bibfnamefont
  {C.}~\bibnamefont {Zimmermann}}, \bibinfo {author} {\bibfnamefont
  {R.}~\bibnamefont {Kleiner}}, \bibinfo {author} {\bibfnamefont
  {D.}~\bibnamefont {Koelle}}, \ and\ \bibinfo {author} {\bibfnamefont
  {J.}~\bibnamefont {Fort\'agh}},\ }\href
  {http://stacks.iop.org/1367-2630/12/i=6/a=065024} {\bibfield  {journal}
  {\bibinfo  {journal} {New J. Phys.}\ }\textbf {\bibinfo {volume} {12}},\
  \bibinfo {pages} {065024} (\bibinfo {year} {2010})}\BibitemShut {NoStop}%
\bibitem [{\citenamefont {Sapina}\ and\ \citenamefont
  {Dahm}(2013)}]{Sapina2013}%
  \BibitemOpen
  \bibfield  {author} {\bibinfo {author} {\bibfnamefont {I.}~\bibnamefont
  {Sapina}}\ and\ \bibinfo {author} {\bibfnamefont {T.}~\bibnamefont {Dahm}},\
  }\href {http://stacks.iop.org/1367-2630/15/i=7/a=073035} {\bibfield
  {journal} {\bibinfo  {journal} {New J. Phys.}\ }\textbf {\bibinfo {volume}
  {15}},\ \bibinfo {pages} {073035} (\bibinfo {year} {2013})}\BibitemShut
  {NoStop}%
\bibitem [{\citenamefont {Naides}\ \emph {et~al.}(2013)\citenamefont {Naides},
  \citenamefont {Turner}, \citenamefont {Lai}, \citenamefont {DiSciacca},\ and\
  \citenamefont {Lev}}]{Naides2014}%
  \BibitemOpen
  \bibfield  {author} {\bibinfo {author} {\bibfnamefont {M.~A.}\ \bibnamefont
  {Naides}}, \bibinfo {author} {\bibfnamefont {R.~W.}\ \bibnamefont {Turner}},
  \bibinfo {author} {\bibfnamefont {R.~A.}\ \bibnamefont {Lai}}, \bibinfo
  {author} {\bibfnamefont {J.~M.}\ \bibnamefont {DiSciacca}}, \ and\ \bibinfo
  {author} {\bibfnamefont {B.~L.}\ \bibnamefont {Lev}},\ }\href {\doibase
  http://dx.doi.org/10.1063/1.4852017} {\bibfield  {journal} {\bibinfo
  {journal} {Appl. Phys. Lett.}\ }\textbf {\bibinfo {volume} {103}},\ \bibinfo
  {eid} {251112} (\bibinfo {year} {2013})}\BibitemShut {NoStop}%
\bibitem [{\citenamefont {Doll}\ and\ \citenamefont
  {N\"abauer}(1961)}]{doll61}%
  \BibitemOpen
  \bibfield  {author} {\bibinfo {author} {\bibfnamefont {R.}~\bibnamefont
  {Doll}}\ and\ \bibinfo {author} {\bibfnamefont {M.}~\bibnamefont
  {N\"abauer}},\ }\href {\doibase 10.1103/PhysRevLett.7.51} {\bibfield
  {journal} {\bibinfo  {journal} {Phys. Rev. Lett.}\ }\textbf {\bibinfo
  {volume} {7}},\ \bibinfo {pages} {51} (\bibinfo {year} {1961})}\BibitemShut
  {NoStop}%
\bibitem [{\citenamefont {Deaver}\ and\ \citenamefont
  {Fairbank}(1961)}]{deaver61}%
  \BibitemOpen
  \bibfield  {author} {\bibinfo {author} {\bibfnamefont {B.~S.}\ \bibnamefont
  {Deaver}}\ and\ \bibinfo {author} {\bibfnamefont {W.~M.}\ \bibnamefont
  {Fairbank}},\ }\href {\doibase 10.1103/PhysRevLett.7.43} {\bibfield
  {journal} {\bibinfo  {journal} {Phys. Rev. Lett.}\ }\textbf {\bibinfo
  {volume} {7}},\ \bibinfo {pages} {43} (\bibinfo {year} {1961})}\BibitemShut
  {NoStop}%
\bibitem [{\citenamefont {Shimizu}\ \emph {et~al.}(2009)\citenamefont
  {Shimizu}, \citenamefont {Hufnagel},\ and\ \citenamefont
  {Mukai}}]{Shimizu2009}%
  \BibitemOpen
  \bibfield  {author} {\bibinfo {author} {\bibfnamefont {F.}~\bibnamefont
  {Shimizu}}, \bibinfo {author} {\bibfnamefont {C.}~\bibnamefont {Hufnagel}}, \
  and\ \bibinfo {author} {\bibfnamefont {T.}~\bibnamefont {Mukai}},\ }\href
  {\doibase 10.1103/PhysRevLett.103.253002} {\bibfield  {journal} {\bibinfo
  {journal} {Phys. Rev. Lett.}\ }\textbf {\bibinfo {volume} {103}},\ \bibinfo
  {pages} {253002} (\bibinfo {year} {2009})}\BibitemShut {NoStop}%
\bibitem [{\citenamefont {Emmert}\ \emph {et~al.}(2009)\citenamefont {Emmert},
  \citenamefont {Lupa\ifmmode~\mbox{\c{s}}\else \c{s}\fi{}cu}, \citenamefont
  {Brune}, \citenamefont {Raimond}, \citenamefont {Haroche},\ and\
  \citenamefont {Nogues}}]{Emmert09}%
  \BibitemOpen
  \bibfield  {author} {\bibinfo {author} {\bibfnamefont {A.}~\bibnamefont
  {Emmert}}, \bibinfo {author} {\bibfnamefont {A.}~\bibnamefont
  {Lupa\ifmmode~\mbox{\c{s}}\else \c{s}\fi{}cu}}, \bibinfo {author}
  {\bibfnamefont {M.}~\bibnamefont {Brune}}, \bibinfo {author} {\bibfnamefont
  {J.-M.}\ \bibnamefont {Raimond}}, \bibinfo {author} {\bibfnamefont
  {S.}~\bibnamefont {Haroche}}, \ and\ \bibinfo {author} {\bibfnamefont
  {G.}~\bibnamefont {Nogues}},\ }\href {\doibase 10.1103/PhysRevA.80.061604}
  {\bibfield  {journal} {\bibinfo  {journal} {Phys. Rev. A}\ }\textbf {\bibinfo
  {volume} {80}},\ \bibinfo {pages} {061604} (\bibinfo {year}
  {2009})}\BibitemShut {NoStop}%
\bibitem [{\citenamefont {M\"uller}\ \emph
  {et~al.}(2010{\natexlab{a}})\citenamefont {M\"uller}, \citenamefont {Zhang},
  \citenamefont {Fermani}, \citenamefont {Chan}, \citenamefont {Wang},
  \citenamefont {Zhang}, \citenamefont {Lim},\ and\ \citenamefont
  {Dumke}}]{Mueller2010a}%
  \BibitemOpen
  \bibfield  {author} {\bibinfo {author} {\bibfnamefont {T.}~\bibnamefont
  {M\"uller}}, \bibinfo {author} {\bibfnamefont {B.}~\bibnamefont {Zhang}},
  \bibinfo {author} {\bibfnamefont {R.}~\bibnamefont {Fermani}}, \bibinfo
  {author} {\bibfnamefont {K.~S.}\ \bibnamefont {Chan}}, \bibinfo {author}
  {\bibfnamefont {Z.~W.}\ \bibnamefont {Wang}}, \bibinfo {author}
  {\bibfnamefont {C.~B.}\ \bibnamefont {Zhang}}, \bibinfo {author}
  {\bibfnamefont {M.~J.}\ \bibnamefont {Lim}}, \ and\ \bibinfo {author}
  {\bibfnamefont {R.}~\bibnamefont {Dumke}},\ }\href
  {http://stacks.iop.org/1367-2630/12/i=4/a=043016} {\bibfield  {journal}
  {\bibinfo  {journal} {New J. Phys.}\ }\textbf {\bibinfo {volume} {12}},\
  \bibinfo {pages} {043016} (\bibinfo {year} {2010}{\natexlab{a}})}\BibitemShut
  {NoStop}%
\bibitem [{\citenamefont {M\"uller}\ \emph
  {et~al.}(2010{\natexlab{b}})\citenamefont {M\"uller}, \citenamefont {Zhang},
  \citenamefont {Fermani}, \citenamefont {Chan}, \citenamefont {Lim},\ and\
  \citenamefont {Dumke}}]{Mueller2010}%
  \BibitemOpen
  \bibfield  {author} {\bibinfo {author} {\bibfnamefont {T.}~\bibnamefont
  {M\"uller}}, \bibinfo {author} {\bibfnamefont {B.}~\bibnamefont {Zhang}},
  \bibinfo {author} {\bibfnamefont {R.}~\bibnamefont {Fermani}}, \bibinfo
  {author} {\bibfnamefont {K.~S.}\ \bibnamefont {Chan}}, \bibinfo {author}
  {\bibfnamefont {M.~J.}\ \bibnamefont {Lim}}, \ and\ \bibinfo {author}
  {\bibfnamefont {R.}~\bibnamefont {Dumke}},\ }\href {\doibase
  10.1103/PhysRevA.81.053624} {\bibfield  {journal} {\bibinfo  {journal} {Phys.
  Rev. A}\ }\textbf {\bibinfo {volume} {81}},\ \bibinfo {pages} {053624}
  (\bibinfo {year} {2010}{\natexlab{b}})}\BibitemShut {NoStop}%
\bibitem [{\citenamefont {Zhang}\ \emph {et~al.}(2012)\citenamefont {Zhang},
  \citenamefont {Siercke}, \citenamefont {Chan}, \citenamefont {Beian},
  \citenamefont {Lim},\ and\ \citenamefont {Dumke}}]{Zhang2012}%
  \BibitemOpen
  \bibfield  {author} {\bibinfo {author} {\bibfnamefont {B.}~\bibnamefont
  {Zhang}}, \bibinfo {author} {\bibfnamefont {M.}~\bibnamefont {Siercke}},
  \bibinfo {author} {\bibfnamefont {K.~S.}\ \bibnamefont {Chan}}, \bibinfo
  {author} {\bibfnamefont {M.}~\bibnamefont {Beian}}, \bibinfo {author}
  {\bibfnamefont {M.~J.}\ \bibnamefont {Lim}}, \ and\ \bibinfo {author}
  {\bibfnamefont {R.}~\bibnamefont {Dumke}},\ }\href {\doibase
  10.1103/PhysRevA.85.013404} {\bibfield  {journal} {\bibinfo  {journal} {Phys.
  Rev. A}\ }\textbf {\bibinfo {volume} {85}},\ \bibinfo {pages} {013404}
  (\bibinfo {year} {2012})}\BibitemShut {NoStop}%
\bibitem [{\citenamefont {Siercke}\ \emph {et~al.}(2012)\citenamefont
  {Siercke}, \citenamefont {Chan}, \citenamefont {Zhang}, \citenamefont
  {Beian}, \citenamefont {Lim},\ and\ \citenamefont {Dumke}}]{Siercke2012}%
  \BibitemOpen
  \bibfield  {author} {\bibinfo {author} {\bibfnamefont {M.}~\bibnamefont
  {Siercke}}, \bibinfo {author} {\bibfnamefont {K.~S.}\ \bibnamefont {Chan}},
  \bibinfo {author} {\bibfnamefont {B.}~\bibnamefont {Zhang}}, \bibinfo
  {author} {\bibfnamefont {M.}~\bibnamefont {Beian}}, \bibinfo {author}
  {\bibfnamefont {M.~J.}\ \bibnamefont {Lim}}, \ and\ \bibinfo {author}
  {\bibfnamefont {R.}~\bibnamefont {Dumke}},\ }\href {\doibase
  10.1103/PhysRevA.85.041403} {\bibfield  {journal} {\bibinfo  {journal} {Phys.
  Rev. A}\ }\textbf {\bibinfo {volume} {85}},\ \bibinfo {pages} {041403}
  (\bibinfo {year} {2012})}\BibitemShut {NoStop}%
\bibitem [{\citenamefont {Scheel}\ \emph {et~al.}(2007)\citenamefont {Scheel},
  \citenamefont {Fermani},\ and\ \citenamefont {Hinds}}]{Scheel2007}%
  \BibitemOpen
  \bibfield  {author} {\bibinfo {author} {\bibfnamefont {S.}~\bibnamefont
  {Scheel}}, \bibinfo {author} {\bibfnamefont {R.}~\bibnamefont {Fermani}}, \
  and\ \bibinfo {author} {\bibfnamefont {E.~A.}\ \bibnamefont {Hinds}},\ }\href
  {\doibase 10.1103/PhysRevA.75.064901} {\bibfield  {journal} {\bibinfo
  {journal} {Phys. Rev. A}\ }\textbf {\bibinfo {volume} {75}},\ \bibinfo
  {pages} {064901} (\bibinfo {year} {2007})}\BibitemShut {NoStop}%
\bibitem [{\citenamefont {Nogues}\ \emph {et~al.}(2009)\citenamefont {Nogues},
  \citenamefont {Roux}, \citenamefont {Nirrengarten}, \citenamefont
  {Lupa{\c{s}}cu}, \citenamefont {Emmert}, \citenamefont {Brune}, \citenamefont
  {Raimond}, \citenamefont {Haroche}, \citenamefont {Pla{\c{c}}ais},\ and\
  \citenamefont {Greffet}}]{Nogues2009}%
  \BibitemOpen
  \bibfield  {author} {\bibinfo {author} {\bibfnamefont {G.}~\bibnamefont
  {Nogues}}, \bibinfo {author} {\bibfnamefont {C.}~\bibnamefont {Roux}},
  \bibinfo {author} {\bibfnamefont {T.}~\bibnamefont {Nirrengarten}}, \bibinfo
  {author} {\bibfnamefont {A.}~\bibnamefont {Lupa{\c{s}}cu}}, \bibinfo {author}
  {\bibfnamefont {A.}~\bibnamefont {Emmert}}, \bibinfo {author} {\bibfnamefont
  {M.}~\bibnamefont {Brune}}, \bibinfo {author} {\bibfnamefont {J.-M.}\
  \bibnamefont {Raimond}}, \bibinfo {author} {\bibfnamefont {S.}~\bibnamefont
  {Haroche}}, \bibinfo {author} {\bibfnamefont {B.}~\bibnamefont
  {Pla{\c{c}}ais}}, \ and\ \bibinfo {author} {\bibfnamefont {J.-J.}\
  \bibnamefont {Greffet}},\ }\href
  {http://stacks.iop.org/0295-5075/87/i=1/a=13002} {\bibfield  {journal}
  {\bibinfo  {journal} {Eur. Phys. Lett.}\ }\textbf {\bibinfo {volume} {87}},\
  \bibinfo {pages} {13002} (\bibinfo {year} {2009})}\BibitemShut {NoStop}%
\bibitem [{\citenamefont {Romero-Isart}\ \emph {et~al.}(2013)\citenamefont
  {Romero-Isart}, \citenamefont {Navau}, \citenamefont {Sanchez}, \citenamefont
  {Zoller},\ and\ \citenamefont {Cirac}}]{Romero2013}%
  \BibitemOpen
  \bibfield  {author} {\bibinfo {author} {\bibfnamefont {O.}~\bibnamefont
  {Romero-Isart}}, \bibinfo {author} {\bibfnamefont {C.}~\bibnamefont {Navau}},
  \bibinfo {author} {\bibfnamefont {A.}~\bibnamefont {Sanchez}}, \bibinfo
  {author} {\bibfnamefont {P.}~\bibnamefont {Zoller}}, \ and\ \bibinfo {author}
  {\bibfnamefont {J.~I.}\ \bibnamefont {Cirac}},\ }\href {\doibase
  10.1103/PhysRevLett.111.145304} {\bibfield  {journal} {\bibinfo  {journal}
  {Phys. Rev. Lett.}\ }\textbf {\bibinfo {volume} {111}},\ \bibinfo {pages}
  {145304} (\bibinfo {year} {2013})}\BibitemShut {NoStop}%
\bibitem [{\citenamefont {Sokolovsky}\ \emph {et~al.}(2014)\citenamefont
  {Sokolovsky}, \citenamefont {Rohrlich},\ and\ \citenamefont
  {Horovitz}}]{Sokolovsky2014}%
  \BibitemOpen
  \bibfield  {author} {\bibinfo {author} {\bibfnamefont {V.}~\bibnamefont
  {Sokolovsky}}, \bibinfo {author} {\bibfnamefont {D.}~\bibnamefont
  {Rohrlich}}, \ and\ \bibinfo {author} {\bibfnamefont {B.}~\bibnamefont
  {Horovitz}},\ }\href {\doibase 10.1103/PhysRevA.89.053422} {\bibfield
  {journal} {\bibinfo  {journal} {Phys. Rev. A}\ }\textbf {\bibinfo {volume}
  {89}},\ \bibinfo {pages} {053422} (\bibinfo {year} {2014})}\BibitemShut
  {NoStop}%
\bibitem [{\citenamefont {Singh}(2009)}]{Singh09}%
  \BibitemOpen
  \bibfield  {author} {\bibinfo {author} {\bibfnamefont {M.}~\bibnamefont
  {Singh}},\ }\href {\doibase 10.1364/OE.17.002600} {\bibfield  {journal}
  {\bibinfo  {journal} {Opt. Express}\ }\textbf {\bibinfo {volume} {17}},\
  \bibinfo {pages} {2600} (\bibinfo {year} {2009})}\BibitemShut {NoStop}%
\bibitem [{\citenamefont {Patton}\ and\ \citenamefont
  {Fischer}(2013{\natexlab{b}})}]{Patton2013a}%
  \BibitemOpen
  \bibfield  {author} {\bibinfo {author} {\bibfnamefont {K.~R.}\ \bibnamefont
  {Patton}}\ and\ \bibinfo {author} {\bibfnamefont {U.~R.}\ \bibnamefont
  {Fischer}},\ }\href {\doibase 10.1103/PhysRevLett.111.240504} {\bibfield
  {journal} {\bibinfo  {journal} {Phys. Rev. Lett.}\ }\textbf {\bibinfo
  {volume} {111}},\ \bibinfo {pages} {240504} (\bibinfo {year}
  {2013}{\natexlab{b}})}\BibitemShut {NoStop}%
\bibitem [{\citenamefont {{Cano}}\ \emph {et~al.}(2011)\citenamefont {{Cano}},
  \citenamefont {{Hattermann}}, \citenamefont {{Kasch}}, \citenamefont
  {{Zimmermann}}, \citenamefont {{Kleiner}}, \citenamefont {{Koelle}},\ and\
  \citenamefont {{Fort{\'a}gh}}}]{Cano2011}%
  \BibitemOpen
  \bibfield  {author} {\bibinfo {author} {\bibfnamefont {D.}~\bibnamefont
  {{Cano}}}, \bibinfo {author} {\bibfnamefont {H.}~\bibnamefont
  {{Hattermann}}}, \bibinfo {author} {\bibfnamefont {B.}~\bibnamefont
  {{Kasch}}}, \bibinfo {author} {\bibfnamefont {C.}~\bibnamefont
  {{Zimmermann}}}, \bibinfo {author} {\bibfnamefont {R.}~\bibnamefont
  {{Kleiner}}}, \bibinfo {author} {\bibfnamefont {D.}~\bibnamefont {{Koelle}}},
  \ and\ \bibinfo {author} {\bibfnamefont {J.}~\bibnamefont {{Fort{\'a}gh}}},\
  }\href {\doibase 10.1140/epjd/e2011-10680-8} {\bibfield  {journal} {\bibinfo
  {journal} {Eur. Phys. J. D}\ }\textbf {\bibinfo {volume} {63}},\ \bibinfo
  {pages} {17} (\bibinfo {year} {2011})}\BibitemShut {NoStop}%
\bibitem [{\citenamefont {London}(1948)}]{London1948}%
  \BibitemOpen
  \bibfield  {author} {\bibinfo {author} {\bibfnamefont {F.}~\bibnamefont
  {London}},\ }\href {\doibase 10.1103/PhysRev.74.562} {\bibfield  {journal}
  {\bibinfo  {journal} {Phys. Rev.}\ }\textbf {\bibinfo {volume} {74}},\
  \bibinfo {pages} {562} (\bibinfo {year} {1948})}\BibitemShut {NoStop}%
\bibitem [{\citenamefont {Brandt}\ and\ \citenamefont
  {Clem}(2004)}]{Brandt2004}%
  \BibitemOpen
  \bibfield  {author} {\bibinfo {author} {\bibfnamefont {E.~H.}\ \bibnamefont
  {Brandt}}\ and\ \bibinfo {author} {\bibfnamefont {J.~R.}\ \bibnamefont
  {Clem}},\ }\href {\doibase 10.1103/PhysRevB.69.184509} {\bibfield  {journal}
  {\bibinfo  {journal} {Phys. Rev. B}\ }\textbf {\bibinfo {volume} {69}},\
  \bibinfo {pages} {184509} (\bibinfo {year} {2004})}\BibitemShut {NoStop}%
\bibitem [{\citenamefont {Cano}\ \emph
  {et~al.}(2008{\natexlab{b}})\citenamefont {Cano}, \citenamefont {Kasch},
  \citenamefont {Hattermann}, \citenamefont {Koelle}, \citenamefont {Kleiner},
  \citenamefont {Zimmermann},\ and\ \citenamefont {Fort\'agh}}]{Cano08}%
  \BibitemOpen
  \bibfield  {author} {\bibinfo {author} {\bibfnamefont {D.}~\bibnamefont
  {Cano}}, \bibinfo {author} {\bibfnamefont {B.}~\bibnamefont {Kasch}},
  \bibinfo {author} {\bibfnamefont {H.}~\bibnamefont {Hattermann}}, \bibinfo
  {author} {\bibfnamefont {D.}~\bibnamefont {Koelle}}, \bibinfo {author}
  {\bibfnamefont {R.}~\bibnamefont {Kleiner}}, \bibinfo {author} {\bibfnamefont
  {C.}~\bibnamefont {Zimmermann}}, \ and\ \bibinfo {author} {\bibfnamefont
  {J.}~\bibnamefont {Fort\'agh}},\ }\href {\doibase 10.1103/PhysRevA.77.063408}
  {\bibfield  {journal} {\bibinfo  {journal} {Phys. Rev. A}\ }\textbf {\bibinfo
  {volume} {77}},\ \bibinfo {pages} {063408} (\bibinfo {year}
  {2008}{\natexlab{b}})}\BibitemShut {NoStop}%
\bibitem [{\citenamefont {Khapaev}\ \emph {et~al.}(2001)\citenamefont
  {Khapaev}, \citenamefont {Kidiyarova-Shevchenko}, \citenamefont {Magnelind},\
  and\ \citenamefont {Kupriyanov}}]{khap}%
  \BibitemOpen
  \bibfield  {author} {\bibinfo {author} {\bibfnamefont {M.}~\bibnamefont
  {Khapaev}}, \bibinfo {author} {\bibfnamefont {A.}~\bibnamefont
  {Kidiyarova-Shevchenko}}, \bibinfo {author} {\bibfnamefont {P.}~\bibnamefont
  {Magnelind}}, \ and\ \bibinfo {author} {\bibfnamefont {M.~Y.}\ \bibnamefont
  {Kupriyanov}},\ }\href {\doibase 10.1109/77.919537} {\bibfield  {journal}
  {\bibinfo  {journal} {IEEE Trans. Appl. Supercond.}\ }\textbf {\bibinfo
  {volume} {11}},\ \bibinfo {pages} {1090} (\bibinfo {year}
  {2001})}\BibitemShut {NoStop}%
\bibitem [{\citenamefont {Smith}\ \emph {et~al.}(2011)\citenamefont {Smith},
  \citenamefont {Aigner}, \citenamefont {Hofferberth}, \citenamefont {Gring},
  \citenamefont {Andersson}, \citenamefont {Wildermuth}, \citenamefont
  {Kr\"{u}ger}, \citenamefont {Schneider}, \citenamefont {Schumm},\ and\
  \citenamefont {Schmiedmayer}}]{Smith2011}%
  \BibitemOpen
  \bibfield  {author} {\bibinfo {author} {\bibfnamefont {D.~A.}\ \bibnamefont
  {Smith}}, \bibinfo {author} {\bibfnamefont {S.}~\bibnamefont {Aigner}},
  \bibinfo {author} {\bibfnamefont {S.}~\bibnamefont {Hofferberth}}, \bibinfo
  {author} {\bibfnamefont {M.}~\bibnamefont {Gring}}, \bibinfo {author}
  {\bibfnamefont {M.}~\bibnamefont {Andersson}}, \bibinfo {author}
  {\bibfnamefont {S.}~\bibnamefont {Wildermuth}}, \bibinfo {author}
  {\bibfnamefont {P.}~\bibnamefont {Kr\"{u}ger}}, \bibinfo {author}
  {\bibfnamefont {S.}~\bibnamefont {Schneider}}, \bibinfo {author}
  {\bibfnamefont {T.}~\bibnamefont {Schumm}}, \ and\ \bibinfo {author}
  {\bibfnamefont {J.}~\bibnamefont {Schmiedmayer}},\ }\href {\doibase
  10.1364/OE.19.008471} {\bibfield  {journal} {\bibinfo  {journal} {Opt.
  Express}\ }\textbf {\bibinfo {volume} {19}},\ \bibinfo {pages} {8471}
  (\bibinfo {year} {2011})}\BibitemShut {NoStop}%
\bibitem [{\citenamefont {{Ketterle}}\ and\ \citenamefont
  {{Druten}}(1996)}]{Ketterle96}%
  \BibitemOpen
  \bibfield  {author} {\bibinfo {author} {\bibfnamefont {W.}~\bibnamefont
  {{Ketterle}}}\ and\ \bibinfo {author} {\bibfnamefont {N.~J.~V.}\ \bibnamefont
  {{Druten}}},\ }\href {\doibase 10.1016/S1049-250X(08)60101-9} {\bibfield
  {journal} {\bibinfo  {journal} {Adv. At. Mol. Opt. Phys.}\ }\textbf {\bibinfo
  {volume} {37}},\ \bibinfo {pages} {181} (\bibinfo {year} {1996})}\BibitemShut
  {NoStop}%
\bibitem [{\citenamefont {{Davis}}\ \emph {et~al.}(1995)\citenamefont
  {{Davis}}, \citenamefont {{Mewes}},\ and\ \citenamefont
  {{Ketterle}}}]{Davis95b}%
  \BibitemOpen
  \bibfield  {author} {\bibinfo {author} {\bibfnamefont {K.~B.}\ \bibnamefont
  {{Davis}}}, \bibinfo {author} {\bibfnamefont {M.-O.}\ \bibnamefont
  {{Mewes}}}, \ and\ \bibinfo {author} {\bibfnamefont {W.}~\bibnamefont
  {{Ketterle}}},\ }\href {\doibase 10.1007/BF01135857} {\bibfield  {journal}
  {\bibinfo  {journal} {Appl. Phys. B}\ }\textbf {\bibinfo {volume} {60}},\
  \bibinfo {pages} {155} (\bibinfo {year} {1995})}\BibitemShut {NoStop}%
\bibitem [{\citenamefont {M\"arkle}\ \emph {et~al.}(2014)\citenamefont
  {M\"arkle}, \citenamefont {Allen}, \citenamefont {Federsel}, \citenamefont
  {Jetter}, \citenamefont {G\"unther}, \citenamefont {Fort\'agh}, \citenamefont
  {Proukakis},\ and\ \citenamefont {Judd}}]{Maerkle2014}%
  \BibitemOpen
  \bibfield  {author} {\bibinfo {author} {\bibfnamefont {J.}~\bibnamefont
  {M\"arkle}}, \bibinfo {author} {\bibfnamefont {A.~J.}\ \bibnamefont {Allen}},
  \bibinfo {author} {\bibfnamefont {P.}~\bibnamefont {Federsel}}, \bibinfo
  {author} {\bibfnamefont {B.}~\bibnamefont {Jetter}}, \bibinfo {author}
  {\bibfnamefont {A.}~\bibnamefont {G\"unther}}, \bibinfo {author}
  {\bibfnamefont {J.}~\bibnamefont {Fort\'agh}}, \bibinfo {author}
  {\bibfnamefont {N.~P.}\ \bibnamefont {Proukakis}}, \ and\ \bibinfo {author}
  {\bibfnamefont {T.~E.}\ \bibnamefont {Judd}},\ }\href {\doibase
  10.1103/PhysRevA.90.023614} {\bibfield  {journal} {\bibinfo  {journal} {Phys.
  Rev. A}\ }\textbf {\bibinfo {volume} {90}},\ \bibinfo {pages} {023614}
  (\bibinfo {year} {2014})}\BibitemShut {NoStop}%
\bibitem [{\citenamefont {Patton}\ and\ \citenamefont
  {Fischer}(2013{\natexlab{c}})}]{Patton2013b}%
  \BibitemOpen
  \bibfield  {author} {\bibinfo {author} {\bibfnamefont {K.~R.}\ \bibnamefont
  {Patton}}\ and\ \bibinfo {author} {\bibfnamefont {U.~R.}\ \bibnamefont
  {Fischer}},\ }\href {http://stacks.iop.org/0295-5075/102/i=2/a=20001}
  {\bibfield  {journal} {\bibinfo  {journal} {Europhys. Lett.}\ }\textbf
  {\bibinfo {volume} {102}},\ \bibinfo {pages} {20001} (\bibinfo {year}
  {2013}{\natexlab{c}})}\BibitemShut {NoStop}%
\bibitem [{\citenamefont {{Zhu}}\ \emph {et~al.}(2011)\citenamefont {{Zhu}},
  \citenamefont {{Saito}}, \citenamefont {{Kemp}}, \citenamefont
  {{Kakuyanagi}}, \citenamefont {{Karimoto}}, \citenamefont {{Nakano}},
  \citenamefont {{Munro}}, \citenamefont {{Tokura}}, \citenamefont {{Everitt}},
  \citenamefont {{Nemoto}}, \citenamefont {{Kasu}}, \citenamefont
  {{Mizuochi}},\ and\ \citenamefont {{Semba}}}]{Zhu2011}%
  \BibitemOpen
  \bibfield  {author} {\bibinfo {author} {\bibfnamefont {X.}~\bibnamefont
  {{Zhu}}}, \bibinfo {author} {\bibfnamefont {S.}~\bibnamefont {{Saito}}},
  \bibinfo {author} {\bibfnamefont {A.}~\bibnamefont {{Kemp}}}, \bibinfo
  {author} {\bibfnamefont {K.}~\bibnamefont {{Kakuyanagi}}}, \bibinfo {author}
  {\bibfnamefont {S.-I.}\ \bibnamefont {{Karimoto}}}, \bibinfo {author}
  {\bibfnamefont {H.}~\bibnamefont {{Nakano}}}, \bibinfo {author}
  {\bibfnamefont {W.~J.}\ \bibnamefont {{Munro}}}, \bibinfo {author}
  {\bibfnamefont {Y.}~\bibnamefont {{Tokura}}}, \bibinfo {author}
  {\bibfnamefont {M.~S.}\ \bibnamefont {{Everitt}}}, \bibinfo {author}
  {\bibfnamefont {K.}~\bibnamefont {{Nemoto}}}, \bibinfo {author}
  {\bibfnamefont {M.}~\bibnamefont {{Kasu}}}, \bibinfo {author} {\bibfnamefont
  {N.}~\bibnamefont {{Mizuochi}}}, \ and\ \bibinfo {author} {\bibfnamefont
  {K.}~\bibnamefont {{Semba}}},\ }\href {\doibase 10.1038/nature10462}
  {\bibfield  {journal} {\bibinfo  {journal} {Nature}\ }\textbf {\bibinfo
  {volume} {478}},\ \bibinfo {pages} {221} (\bibinfo {year}
  {2011})}\BibitemShut {NoStop}%
\end{thebibliography}
\end{document}